\documentstyle[12pt]{article}
\def\gsim{\stackrel{>}{\sim}}
\def\lsim{\stackrel{<}{\sim}}
\def\beq{\begin{equation}}
\def\eeq{\end{equation}}

\begin{document}
\begin{flushright}
VAND--TH-94--17\\
July 1994
\end{flushright}
\begin{center}
\Large
{\bf The Higgs Mass as the Discriminator of Electroweak Models}
\end{center}
\normalsize
\bigskip
\begin{center}
{\large Marco A.  D\'\i az, Tonnis A. ter Veldhuis} and
{\large Thomas J. Weiler}\\
{\sl Department of Physics \& Astronomy\\
Vanderbilt University\\
Nashville, TN 37235, USA}\\
\vspace{2cm}
\end{center}
\centerline{ABSTRACT}

In the Minimal Supersymmetric Model (MSSM) and the Next to Minimal
Supersymmetric Model [(M+1)SSM], an upper bound on
the lightest higgs mass can be calculated. On the other hand, vacuum
stability implies a lower limit on the mass of the higgs boson in the
Standard Model (SM). We find that a gap exists for
$m_t \stackrel{>}{\sim} 165$ GeV between
the SM and both the MSSM and the (M+1)SSM bounds. Thus, if the new top
quark mass measurement by CDF remains valid, a first measurement of the
higgs mass will serve to exclude either the SM or the MSSM/(M+1)SSM higgs
sectors. In addition, we discuss  Supersymmetric Grand Unified Theories, other
extentions of the SM, the discovery potential of the lightest higgs, and
the assumptions on which our conclusions are based.

PACS numbers: 12.60Fr, 12.60Jv, 12.15Lk, 14.80Cp. 14.80Bn
\vfill\eject

The Achilles' heel of the Standard Model (SM) is the electroweak
symmetry--breaking sector.  The simplest and most motivated
possibilities for this sector are the single higgs doublet of the
minimal SM, and the two higgs doublet sector of the Minimal Supersymmetric
Standard Model (MSSM).
After roughly twenty years of experimental efforts to expose the origin
of broken weak symmetry, not a single clue has been found.
In particular, the simple SM and MSSM possibilities remain viable
but unconfirmed.  Recently, hope has risen that a new window to the
symmetry--breaking sector may have been found:
the CDF experiment at Fermilab has
announced \cite{CDF} the probable discovery of the top
quark with mass at $174\pm16$ GeV
(very consistent with the SM prediction of
$m_t=164\pm25$ GeV inferred from precision electroweak data \cite{ewgroup}).
This range of $m_t$ values encompasses the electroweak symmetry--breaking
scale,
defined by the vacuum expectation value (vev)
$<0|\Phi|0>=v_{SM}/\sqrt{2}=175$ GeV of the complex higgs field $\Phi$.
The fact that the central CDF value is nearly
identical to the $\Phi$ vev is intriguing, and presumably coincidental.
The fact that the eventual true value of $m_t$ will be comparable to
the symmetry--breaking scale is fortuitous, for it suggests that the top quark
may communicate the secrets of the symmetry--breaking to us either through
top properties, or through large quantum corrections to classical physics.
In the SM and in supersymmetric (susy) models the main
uncertainty in radiative corrections is the value of the top mass.
If the CDF announcement is confirmed, this main uncertainty is eliminated.
One observation \cite{kp} which we quantify in this Letter
is the following:
inputing the CDF value for the top mass into quantum loop
corrections for the symmetry-- breaking higgs sector leads to mutually
exclusive, reliable bounds on the SM higgs mass and
on the lightest MSSM higgs mass.  From
this we infer that
{\it if the CDF value for $m_t$ is
verified in the 1994--95 data run, then the first higgs mass measurement
will rule out one of the two main contenders (SM vs.\ MSSM)
for \underline{the} electroweak theory, independent of any other measurement.}

Another point deserves emphasis.
It is known that the Feynman rules connecting the lightest higgs in the MSSM
to ordinary matter become exactly the SM Feynman rules, in the limit
where the ``other'' higgs masses (these are $m_A, m_H$, and $m_{H^{\pm}}$,
found in any two-higgs-doublet models) are taken to infinity \cite{HHG}.
When the masses are taken
large compared to $M_Z$, of the order of
a TeV for example, the lightest MSSM higgs behaves very much
like the SM higgs in its production channels and decay modes \cite{cpr1}.
Furthermore, the mass of the lightest MSSM higgs rises toward its upper bound
as the ``other'' higgs masses are increased.
Thus, for masses in the region where the SM lower bound
and the MSSM upper bound overlap,
the SM higgs and the lightest MSSM higgs may not be distinguishable by
branching ratio or width measurements.
Only if the two bounds are separated by a gap is this ambiguity avoided.
Thus, {\it there may be no discernible difference between the lightest
MSSM higgs and the SM higgs, except for their allowed mass values.}
The gap develops with increasing $m_t$ because
the MSSM higgs self--coupling is constrained,
a vestige of the underlying supersymmetry,
and thus requires the MSSM higgs to be light,
whereas SM vacuum stability requires the SM higgs to be heavy.
We demonstrate the onset of the mass gap in Fig. 1.

Recently it has been shown that when the newly reported
value of the top mass is input
into the effective potential for the SM higgs field, the
broken--symmetry potential minimum is stable
\footnote{
If the universe is allowed to reside in an unstable minimum,
then a similar, but slightly weaker (by $\lsim 5$ GeV for
heavy $m_t$ \cite{sher}) bound results \cite{arnold}.
}
only if the SM higgs mass satisfies the lower bound constraint \cite{sher}:
\begin{equation}
m_H > 132+2.2(m_t-170)-4.5(\frac{\alpha_s - 0.117}{0.007}),    \label{eq:sher}
\end{equation}
valid for a top mass in the range 160 to 190 GeV. In this equation,
mass units are in GeV, and $\alpha_s$ is the strong coupling constant
at the scale of the $Z$ mass.  This equation is the result of RGE--improved
two--loop calculations, and includes radiative corrections to the higgs and
top masses.  It is reliable
\footnote
{If we use  the generous
value $\alpha_s=0.129$, the lower bound on the SM higgs mass decreases
by about 8 GeV for $m_t>160$ GeV.
A decrease of even this magnitude in the SM lower bound
is compensated by the decrease in the MSSM upper bound due to
two-loop contributions not included in our calculations, but discussed
in the text.
},
and accurate to 1 GeV in the top mass, and
2 GeV in the higgs mass \cite{sher}.

It has been known for some time \cite{reviews}
that the SM lower bound
rises rapidly as the value of the top mass increases through $M_Z$; below $M_Z$
the bound is of order of the Linde--Weinberg value, $\sim 7$ GeV \cite{LW}.
So what is new here
is the inference from the large reported value for $m_t$ that the SM higgs
lower mass bound dramatically exceeds 100 GeV!
The D0 collaboration has used its
nonobservation of top candidates to report a
95\% confidence level lower bound on the top mass of 131 GeV \cite{D0}.
Thus, the D0 lower bound, and the CDF mass value including $1\sigma$
allowances are, respectively, 131, 158, 174, and 190 GeV.
Inputing these top mass values into Eq.~(\ref{eq:sher}) and
the equivalent for the lower range of $m_t$ \cite{sher}
with $\alpha_s = 0.117$ then yields SM higgs mass lower bounds of
60, 106, 140, and 176 GeV, respectively.

This lower limit on the SM higgs from the vacuum stability argument
is a significant
phenomenological constraint, and it rises linearly with
$m_t$, for $m_t \stackrel{>}{\sim} 100$ GeV.
On the other hand, the upper limit on the lightest MSSM higgs rises
quadratically with $m_t$, also for $m_t \stackrel{>}{\sim} 100$ GeV.
In fact,
{\it the radiatively corrected observable
most sensitive to the value
of the top mass is the mass of this
lightest higgs particle in susy models {\rm \cite{mt4}}}:
for large top mass, the top and scalar--top ($\tilde{t}$) loops dominate
all other loop corrections, and
{\it the light higgs mass--squared grows as }
$m_t^4 \ln(m_{\tilde{t}}/m_t)$
\footnote{
Note that the correction grows logarithmically
as $m_{\tilde{t}}$ gets heavy, rather than decoupling!
For heavy $m_{\tilde{t}}$ the large logarithms can be summed to all orders
in perturbation theory using renormalization group techniques.
Interestingly, the effect is to \underline{lower}
the MSSM upper bound \cite{haberralf}.
}
{}.
Thus, for very heavy $m_t$, the two bounds will inevitably overlap.
Also, for relatively light $m_t$ the bounds may overlap; e.g.\ we
have just seen that the SM lower bound is 60 GeV for $m_t = 131$ GeV,
whereas for large or small $\tan\beta$ the MSSM upper bound is at least
the Z mass.  However, for $m_t$ around the value reported
by the CDF collaboration, we demonstrate by careful calculation that
there is a gap between the SM higgs mass
lower bound and the MSSM upper bound.
Thus, the first measurement of the lightest higgs mass
will serve to exclude either the SM higgs sector, or the MSSM higgs sector!

Since vacuum stability of the SM first breaks down for scalar field
fluctuations on the order of $10^6 - 10^{10}$ GeV \cite{sher},
an implicit assumption in this
SM bound is no new physics below $10^{10}$ GeV.
In particular, the stability bound, calculated with perturbation theory,
is not valid if there is a non--perturbatively large value for
the higgs self--coupling $\lambda$
below $\sim 10^{10}$ GeV.
However, if there is a non--perturbatively large value for $\lambda$
below $10^{10}$ GeV, then there will be a Landau pole near or below
$10^{10}$ GeV, which in turn implies a
{\it triviality \underline{lower} bound} on the SM higgs mass of
about 210 GeV.
A derivation and discussion of this triviality lower bound, as well as further
details related to the matters of this Letter, are given in
\cite{DVW}.
An immediate consequence is:
{\it assuming no new fields with mass scales below $10^{10}$ GeV,
either the perturbative stability bound is valid for the SM higgs,
or the non--perturbative triviality lower bound is valid.}
The stability bound is the less restrictive, and we assume it in the
subsequent sections of this paper.

Turning now to the MSSM model,
we calculate the one-loop corrected lightest MSSM higgs mass,
$m_h$ \cite{marco},
including the full one--loop corrections from the
top/bottom quarks and squarks,
and the leading--log corrections from
the remaining fields (charginos, neutralinos,
gauge bosons, and higgs bosons).
As advertised earlier,
the mass corrections are sizeable
\footnote{
Notice that the lightest higgs mass,
bounded at tree--level by $m_h\leq |\cos(2\beta)| \, M_Z$,
vanishes at tree level if $\tan\beta=1$.
However, radiative
corrections so strongly modify this tree level prediction that
the $\tan\beta=1$ scenario remains viable \cite{diazhaberii}.
}.
Recently, full one--loop corrections from all particles \cite{1loop}
have been calculated.
Since the dominant corrections are due to the heavy quarks and squarks,
full one--loop corrections from charginos, neutralinos, gauge and higgs bosons
are well approximated by their leading logarithm terms used here.
Two--loop corrections have recently been calculated also \cite{2loop},
for the limit where the ratio of the vacuum expectation
values $\tan\beta \rightarrow \infty$.
Keeping only the leading $m_t$ terms, these corrections have been extrapolated
to all $\tan\beta$.  The graphical result in ref. \cite{2loop}
shows a \underline{lowering} of the MSSM upper bound by several GeV
\footnote
{In ref.\ \cite{eq} were found small two-loop contributions of the
order $m_t^6$; however, the QCD two-loop contributions found in
ref. \cite{2loop} are
of order $\alpha_s^2m_t^4$, and dominate the previous ones.
The net effect is to lower the higgs mass bound.
}.
{}From this work \cite {2loop}, we estimate the gap to be wider by several GeV
than the one--loop separation we show in our figure.
This widening further enables a higgs mass measurement to distinguish
the SM and MSSM models.

The lightest higgs mass as a
function of $\tan\beta$ is shown in Fig. 1.
For the case $\tan\beta\sim 1$, the SM lower bound and the MSSM upper bound
are already non--overlapping at $m_t=131$ GeV\@.
However, for larger $\tan\beta$ values, the overlap persists until
$m_t \stackrel{>}{\sim} 165$ GeV\@.
For the preferred CDF value of $m_t=174$GeV, the gap is present for all
$\tan\beta$, allowing discrimination between the SM and the MSSM based
on the lightest higgs mass alone.  At $m_t=190$ GeV the gap is still widening,
showing no signs of the eventual gap--closure at still higher $m_t$.
It is reassuring that the upper bounds in the region
of acceptable $\tan \beta$  are similar
for small and large squark mixing.

The MSSM can be extended in a
straightforward fashion by adding an $SU(2)$ singlet $S$ with vanishing
hypercharge to the theory \cite{Ell}.
As a consequence, this (M+1)SSM model contains an
additional scalar, pseudoscalar, and neutralino.
A tree--level analysis of the eigenvalues
of the scalar mass matrix yields an upper bound on the mass of the
lightest higgs boson:
\begin{equation}
m_h^2 \leq M_Z^2 \left\{ \cos^2 2 \beta +2 \frac{\lambda^2}{g_1^2 + g_2^2}
\sin^2 2 \beta \right\}.
\end{equation}
The new higgs self coupling
$\lambda$ is {\it a priori} free, and so the second
term may \underline{considerably}
weaken the upper bound \cite{Vel,nonVel}.
However, there are two
cases where the bound will suffer only a minor adjustment.
The first is the large $\tan\beta$ scenario, where $\cos^2 2\beta$ is
necessarily $\gg \sin^2 2\beta$.  The second is
when the theory is embedded into a GUT;
even if $\lambda$ assumes a high value at the GUT scale,
the nature of the
renormalization group equations is such that
its evolved value at the susy--breaking scale
is a rather low, pseudo--fixed point.
Under the assumption that all coupling constants remain perturbative
up to the GUT scale, it is therefore possible to calculate a maximum value
for the mass of the lightest higgs boson \cite{Vel,nonVel}.
The higgs mass upper bound depends on the value
of the top yukawa at the GUT scale through the renormalization group
equations.

In Fig. 1 we show the maximum value of the
higgs boson mass as a function of $\tan \beta$ for
the chosen values of the top quark mass $m_t$.
The bounds are quite insensitive to the choice of $M_{SUSY}$,
increasing very slowly
as $M_{SUSY}$ increases \cite{Vel}.
It is revealed that
for low values of the top quark mass ($\sim M_Z$), the mass upper bound on the
higgs boson in the
(M+1)SSM will be substantially higher than in the MSSM at
$\tan\beta \lsim$ a few.
However, for a larger top quark mass the difference
between the MSSM and (M+1)SSM upper bounds diminishes.
There is a minimum allowed $\tan\beta$ in the (M+1)SSM,
implied by the top yukawa pseudo--fixed--point.
The minimum rises with $m_t$, and
is evident in the figure.
The (M+1)SSM and MSSM bounds are
very similar at $\tan\beta\gsim 6$ (the only viable region in
the (M+1)SSM model for $m_t$ at or above the CDF value).
Since the (M+1)SSM model was originally constructed to test the robustness of
the MSSM, it is gratifying that the two models show a very similar upper bound.

The results for more complicated extensions of the minimal susy model tend
to be similar \cite{Kan}. In general, the mass of the lightest higgs boson
at tree level is limited by $M_Z$ times a factor proportional to the
dimensionless coupling constants in the higgs sector.
The requirement
of perturbative unification restricts the  value of these coupling
constants at the electroweak scale, and the maximum value of the lightest higgs
boson mass is therefore never much larger than $M_Z$.

We have seen that the SM, MSSM, and the (M+1)SSM electroweak models can be
disfavored or ruled out by a measurement of $m_h$;
and that a ``forbidden'' mass gap exists for these models if
$m_t\gsim 165$ GeV.
A summary of these mass bounds
\footnote{
In constructing this table, we have taken the values of the MSSM higgs
upper bound without considering the region $\tan\beta\ll 1$.
Large radiative corrections
appear at $\tan\beta\ll 1$ because the value of the top yukawa coupling
is extrapolated beyond what is perturbatively valid;
the result is suspect.
The argument of perturbative validity argues against
this small $\tan\beta$ region \cite{lowtbeta}.
}
is provided in table \ref{tbl},
for four possible $m_t$ values.
\begin{table}
\begin{center}
\begin{tabular}{|ll|cccc|}
\hline
$m_t$    &             & 131 & 158 & 174 & 190 \\
\hline
SM       &  $m_h\;>$   & 60  & 106 & 140 & 176 \\
MSSM     &  $m_h\;<$   & 104  & 119 & 130 & 143 \\
(M+1)SSM &  $m_h\;<$   & 136  & 129 & 128 & 133 \\
\hline
\end{tabular}
\end{center}
\caption{The SM lower bound and the MSSM and (M+1)SSM upper
bounds on the higgs boson mass $m_h$ for various values of the top quark mass
$m_t$.} \label{tbl}
\end{table}
However, some other models do not tightly constrain the lightest higgs
mass. Examples of such models are the SM without a desert
\cite{lindner},
non--minimal SUSY with unconstrained higgs self--coupling, and low energy
effective models of strongly coupled theories
\cite{tonnis}.
These models cannot be ruled out by a single higgs mass measurement.

Many supersymmetric grand--unified theories (susy GUTs)
reduce at low energies to the MSSM
with additional constraints on the parameters.
Accordingly, the upper limit
on $m_h$ in such susy GUTs is
in general \underline{more} \underline{restrictive}
than the bound presented here.
The assumption of the pseudo--fixed--point solution for the top mass is
an attractive example because
the apparent CDF top mass value is within the estimated range of the
pseudo--fixed--point \cite{candm}.
With the additional assumptions
that the electroweak symmetry is radiatively broken
and that the low energy MSSM spectrum is defined by a small number of
parameters at the GUT scale,
two compact, disparate allowed ranges for $\tan\beta$ emerge:
$1.0\leq \tan\beta \leq 1.4$ \cite{bbk},
and a large $\tan\beta$ solution $\sim m_t/m_b$, disfavored by proton stability
arguments \cite{nath}.
In fact, a highly constrained
low $\tan\beta$ region $\sim 1$ and high $\tan\beta$ region
$\gsim$ 40--70
also emerge when
bottom--$\tau$ yukawa unification at the GUT scale is imposed on
the radiatively broken model \cite{susygut}.
The small $\tan\beta$ restriction results when the
top mass, but not the bottom mass, is assigned to its pseudo--fixed--point.
Resulting mass bounds in the literature are basically our bound
in Fig. 1 for $\tan\beta\sim 1-3$.
The net effect of the yukawa--unification constraint in susy GUTs is
necessarily to widen the mass gap
between the light higgs MSSM and the heavier higgs SM,
thus strengthening the potential for experiment to distinguish the models.

We end with conclusions on detectability of the lightest higgs
\cite{gunion,mrenna}.
If $m_t \sim 131$ GeV, then
the SM higgs mass lower bound from vacuum stability
is 60 GeV; a SM mass up to (80,105) GeV is
detectable at (LEP178,LEP200), and a SM mass up to 130 GeV is detectable
at a High Luminosity Di Tevatron (HLDT) \cite{mrenna};
the MSSM $h^0$ is
certainly detectable at LEP178 for $\tan\beta \sim$ 1--2, and
certainly detectable at LEP200 for all $\tan\beta$.
If $m_t \sim 174$ GeV, then
the SM higgs is above 140 GeV, out of reach for LEPII and the HLDT;
the MSSM higgs is certainly detectable at LEP200 if $\tan\beta \sim$ 1--2.
Conclusions for $m_t=$ 158 and 190 geV can be inferred after reference to
Table 1.
It is interesting that the $h^0$ mass range is most accessible to experiment
if $\tan\beta \sim$ 1--3, just the parameter range favored by susy GUTs.

We repeat that the lightest MSSM higgs is guaranteed detectable at LEP230;
and that the lightest (M+1)SSM higgs and MSSM higgs
are guaranteed detectable at a NLC300 and at the LHC.
Since there is no lower bound on the lightest MSSM higgs mass other than the
experimental bound,
the MSSM $h^0$ is possibly detectable even at LEP178 for all $\tan\beta$,
but there is no guarantee.
The SM higgs is guaranteed detectable only at the LHC;
if $m_t \sim 174$ GeV, then the SM higgs will not be produced until the LHC
or NLC is available.
Thus, one simple conclusion is that LEPII has a tremendous potential to
distinguish MSSM and (M+1)SSM symmetry
breaking from SM symmetry breaking.

In conclusion,
we have shown that for a top quark mass $\sim 174$ GeV, as reported by CDF,
a gap exists between the SM higgs mass ($\gsim 140$ GeV) and the lightest
MSSM higgs mass ($\lsim 130$ GeV).  Thus, the first higgs mass measurement
will eliminate one of these popular models.
Most of the MSSM mass range is accessible to LEPII.
If a higgs is discovered
at LEPII, the SM higgs sector is ruled out.
We remind the reader that our conclusions regarding the SM assume
a desert up to (at least) $10^{10}$ GeV.
For the (M+1)SSM with the assumption of perturbative unification,
conclusions remain the same as for the MSSM.
\vspace{1.0cm}
\\
{\bf Acknowledgements:}\\
This work was supported in part by the U.S. Department of Energy grant
no.\ DE-FG05-85ER40226, and the Texas National Research Laboratory Commission
grant no.\ RGFY93--303.
\vfill\eject

\vfill\eject
\noindent
{\bf Figure Caption:}\\
{\bf Fig. 1.}
Higgs mass as a function of $\tan\beta$ for two different
values of the top quark mass (a) $m_t=174$ and (b) $m_t=
131$ GeV. We show the SM lower bound (dotdash), the
(M+1)SSM upper bound (solid) with a GUT scale given by
$10^{16}$ GeV, and the MSSM upper bound
(dashes). In the MSSM case curves are shown for two
different choices of the squark mixing parameters: no
squark mixing ($\mu=A=0$) and maximal mixing ($\mu=A=1$
TeV). The first choice is the one where the higgs mass
approaches asymptotically to a constant as $\tan\beta$
increases. In all cases, every superparticle and higgs
beyond the lightest are assumed to have a mass of the
order of 1 TeV.
\\
\end{document}